\documentclass[11pt,a4paper]{article}
\usepackage{float} % place les figure au bon endroit
\usepackage{bm}
\usepackage{amsmath}
\usepackage{amsfonts}
\usepackage{amssymb}
\usepackage{graphicx}
\usepackage[left=1.8cm,right=1.8cm,top=1.8cm,bottom=1.8cm]{geometry}

\usepackage{color}
\usepackage{soul}

\usepackage[utf8]{inputenc}
\usepackage[T1]{fontenc}
\usepackage{lmodern}
\usepackage{authblk}
\usepackage{gensymb}
\usepackage{lineno}

\usepackage{titlesec}
\titleformat*{\section}{\bfseries\Large}
\titleformat*{\subsection}{\bfseries \normalsize}
\titleformat*{\paragraph}{\bfseries \normalsize}

\newcommand{\eq}[1]{Eq.~\eqref{#1}}
\newcommand{\fig}[1]{Fig.~\ref{#1}}

\setstcolor{red}

\title{Correlated spontaneous emission of fluorescent emitters mediated by single plasmons}

\author[1]{Dorian~Bouchet}
\author[2]{Emmanuel~Lhuillier}
\author[3]{Sandrine~Ithurria}
\author[4]{Angelo Gulinatti} 
\author[4]{Ivan Rech}
\author[1]{Rémi~Carminati}
\author[1]{Yannick~De~Wilde}
\author[1]{Valentina~Krachmalnicoff*}

\affil[1]{\footnotesize ESPCI ParisTech, PSL Research University, CNRS, Institut Langevin, 1 rue Jussieu, F-75005 Paris, France}
\affil[2]{Institut des NanoSciences de Paris, UPMC-UMR CNRS 7588, 4 place Jussieu, 75252 Paris CEDEX 05, France}
\affil[3]{ESPCI ParisTech, PSL Research University, CNRS, Sorbonne Universités, UPMC Univ. Paris 6; LPEM, 10 rue Vauquelin, F-75231 Paris Cedex 5, France}
\affil[4]{Politecnico di Milano, Dipartimento di Elettronica, Informazione e Bioingegneria, Piazza da Vinci 32, 20133 Milano, Italy}

\begin{document}

\maketitle

\bigskip

Manipulating the spontaneous emission of a fluorescent emitter can be achieved by placing the emitter in a nanostructured environment. A privileged spot is occupied by plasmonic structures that provide a strong confinement of the electromagnetic field, which results in an enhancement of the emitter-environment interaction. While plasmonic nanostructures have been widely exploited to control the emission properties of single photon emitters \cite{akimov_generation_2007},\cite{chang_single-photon_2007},\cite{kolesov_waveparticle_2009},\cite{Greffet_Maitre_NanoLetters2013}, performing the coupling between quantum emitters with plasmons poses a huge challenge \cite{tame_quantum_2013}. In this Letter we report on a first crucial step towards this goal by the observation of correlated emission between a single CdSe/CdS/ZnS quantum dot exhibiting single photon statistics and a fluorescent nanobead located micrometers apart. This is accomplished by coupling both emitters to a silver nanowire. Single-plasmons are created on the latter from the quantum dot, and transfer energy to excite in turn the fluorescent nanobead.  
\begin{center}
\line(1,0){200}
\end{center}
\bigskip

The observation of the correlated emission of two quantum emitters has implications covering a wide panel of fields. It constitutes the keystone for more elaborate experiments crucial for quantum technologies such as producing entangled states, or realizing interactions among a small ensemble of quantum emitters within the same plasmonic mode along a bus on an integrated device \cite{Faez_Sandoghdar_PRL2014}, or studying cooperative emission phenomena like superradiance or subradiance \cite{martin-cano_resonance_2010}. The communication between two emitters via the exchange of photons propagating in free space was demonstrated so far at temperatures near absolute zero, taking advantage that in that case a molecule with a transition wavelength $\lambda$ resonant with an incoming photon has a extinction cross section on the order of ${\lambda^2}/2$ , which is comparable to the area of a beam focused with a high numerical aperture objective \cite{Rezus_Sandoghdar2012}. Here, we go one step further and we address the problem of the communication between two distant non resonant emitters at room temperature on a nanostructured waveguide. While we use a quantum emitter as a source, we compensate for the low extinction cross section of nonresonant molecules at room temperature using a one dimensional plasmonic waveguide made of silver and a receiver which consists in a fluorescent nanobead containing a large amount of fluorescent molecules in near-field coupling with the plasmonic waveguide.    

Silver nanowires have been suggested as broadband waveguides, supporting surface plasmons which propagate over distances of several micrometers along the wire axis \cite{ditlbacher_silver_2005}. The strong electromagnetic field confinement which characterizes surface plasmons leads to large enhancements of the spontaneous emission rate of emitters near-field coupled to silver nanowires, which go along with an efficient coupling into the guided surface plasmon mode \cite{chang_quantum_2006}. It has also been shown that the coupling of single photon emitters, e.g. semiconductor quantum dots (QD) or nitrogen-vacancy defects, to a silver nanowire, generates single surface plasmons exhibiting properties similar to those of single photons \cite{akimov_generation_2007}\cite{kolesov_waveparticle_2009}. Several theoretical studies highlighted the potential of coupling quantum emitters to surface plasmons \cite{dzsotjan_quantum_2010}\cite{martin-cano_resonance_2010}. In the meantime, experimental observations of plasmon-assisted energy transfer have so far involved ensembles of fluorescent emitters \cite{andrew_energy_2004}\cite{bouchet_long-range_2016}\cite{de_torres_coupling_2016}. 

\bigskip
To perform our experiment, we take benefit of the strong interaction between a single QD and a silver nanowire, as revealed by the measurement of a high Purcell factor $\Gamma/\Gamma_0 \sim 20$ ($\Gamma$ and $\Gamma_0$ being the spontaneous decay rates in the presence and in the absence of the nanowire, respectively), and realize long-range energy transfer between this single nanocrystal and a fluorescent nanobead, located on the same nanowire 8.7 $\mu$m apart. Using time-resolved measurements, we provide a comprehensive analysis of the relation between photon emission from the QD and the fluorescent nanobead. For the sample preparation, we disperse a dilute solution of chemically synthesized CdSe/CdS/ZnS quantum dots \cite{talapin_cdse/cds/zns_2004} and polystyrene fluorescent beads (mean diameter $\sim$176~nm) on a glass coverslip presenting several silver nanowires (see Methods). Figure 1a presents the optical setup of the experiment, based on an inverted fluorescence microscope combined with time-resolved single photon detection. 
\begin{figure}[ht]
\centering
\includegraphics[scale = 0.92]{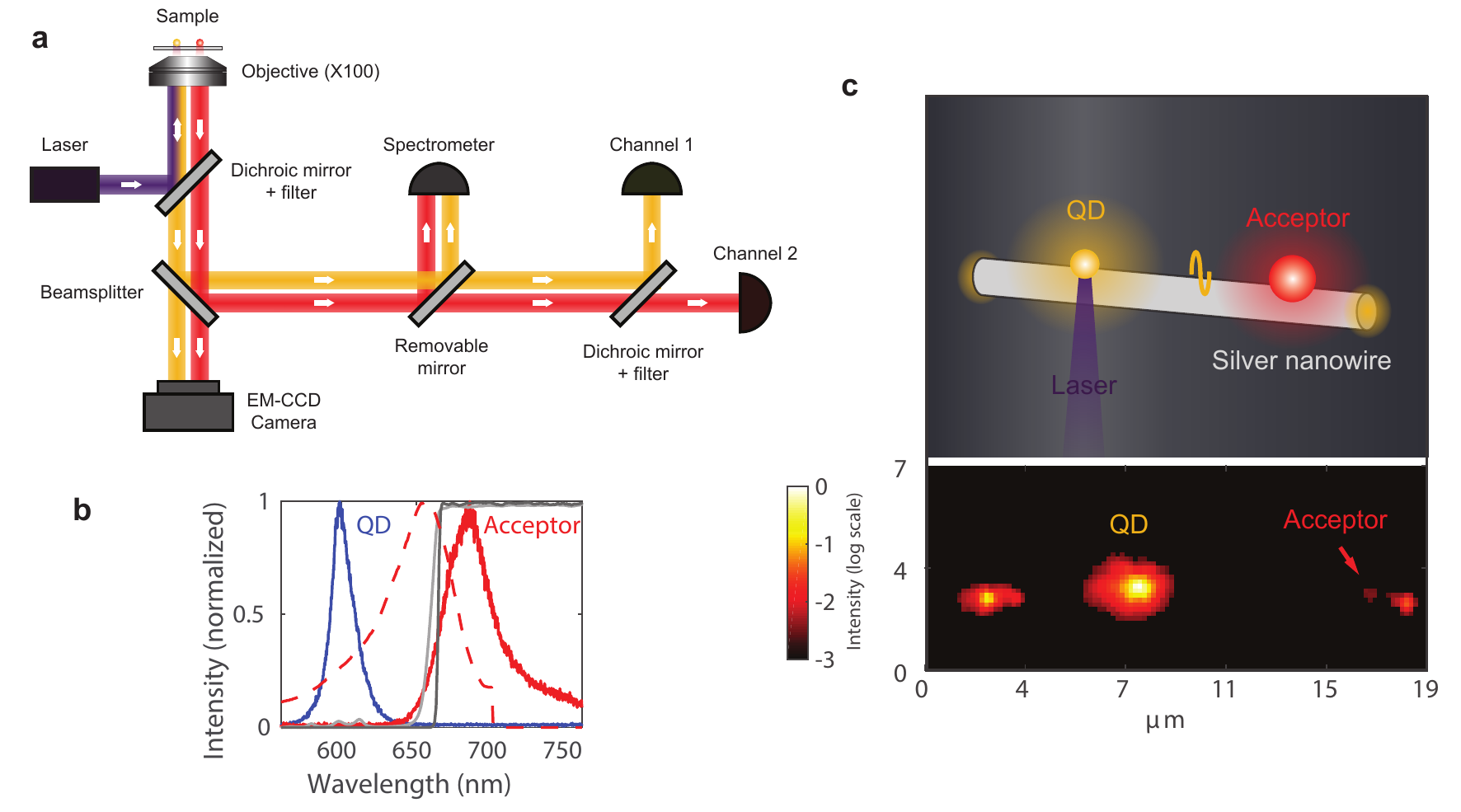} 
\caption{Description of the experiment. \newline \hspace{\textwidth} \small \textbf{a,} Optical setup used for imaging, spectroscopy, and time-resolved measurements of a QD (donor) and a fluorescent bead (acceptor) coupled to a silver nanowire. \textbf{b,} Normalized emission spectra of the QD (in blue) and the acceptor bead (in red). The absorption spectrum of the bead is the dashed red curve. Transmittance of the dichroic mirror and of the long-pass filter are represented in light grey and grey, respectively. \textbf{c,} Top: Artist view of the experiment. Bottom: EM-CCD image taken during the experiment, while the laser is focused on the QD. Bright spots are measured at the position of both emitters as well as both ends of the nanowire.}
\label{f1}
\end{figure}
A pulsed laser diode emitting at $\lambda = 405$~nm and at a repetition rate of 40~MHz is used to excite the emitters. An oil immersion microscope objective (x100 magnification, NA=1.4) located below the sample ensures both illumination and an efficient collection of fluorescence photons. QDs exhibit large absorption at $\lambda = 405$~nm while acceptor beads have a maximum absorption around $\lambda = 640$~nm, limiting direct excitation of the acceptor beads by the laser.  %Note that these last have an absorption (??) cross section estimated at $\sigma=??$ at room temperature, which is one (??) order of magnitude larger than that of a resonant single molecule at cryogenic temperature \cite{Rezus_Sandoghdar2012}. 
Using wide-field images captured with an EM-CCD camera, we select a single QD and a single acceptor bead, both of them located in the near field of a 16~$\mu$m long silver nanowire. This configuration is represented in \fig{f1}c (top). Focusing the laser onto the QD, we observe plasmon scattering from the extremities of the nanowire (\fig{f1}c, bottom). A bright spot also appears at the position of the acceptor. From this image, we determine a distance $d$ of 8.7~$\mu$m between the QD and the acceptor bead (see Methods). In the same time, we perform time resolved measurements using two single photon avalanche diodes (SPADs), using a dichroic mirror and a long-pass filter to split the incident light. The SPADs are placed in such a way that channel~1 measures the photon emission from the QD and channel~2 measures the emission from the acceptor bead. Emission spectra (shown in \fig{f1}b) are measured with a fibered spectrometer. The absence of overlap between both emission spectra ensures that no photon emitted by the QD is detected on channel 2.

\bigskip

In order to prove that the acceptor is excited via plasmon-mediated energy transfer, we need to characterize the decay histogram of the QD and the acceptor excited independently. Figure 2a presents the time dependence of the fluorescence counts of the QD, binned at 1~ms.
\begin{figure}[ht]
\centering
\includegraphics[scale = 1.06]{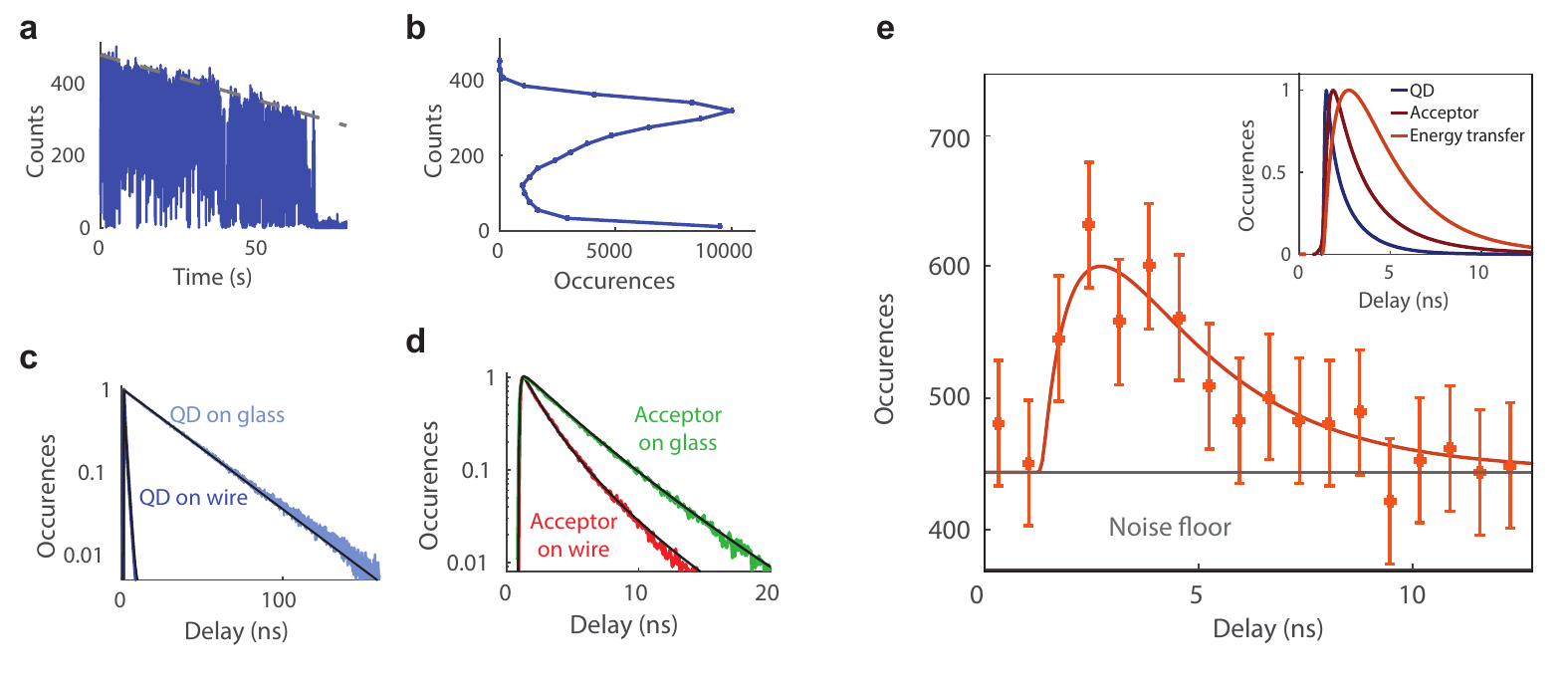} 
\caption{Fluorescence time trace and decay rate analysis. \newline \hspace{\textwidth} \small \textbf{a,} Time dependence of the fluorescence counts of the QD binned at 1~ms (in blue). The drift of the experiment is represented by the dashed line. \textbf{b,} Distribution of detected events, showing a clear distinction between the bright and the dark state of the QD. \textbf{c,} Fluorescence decay histograms of a single QD on glass (light blue) and on the nanowire (dark blue). \textbf{d,} Fluorescence decay histogram for the acceptor on glass and on wire, when excited via a plasmon. \textbf{e,} Measured decay statistics of the acceptor together with the model function (in orange). The width of error bars is equal to the square root of the total number of detected events. Inset: Fit functions of the decay histograms of the QD (blue curve) and the acceptor (red curve) excited independently and their convolution (orange curve).}
\label{f2}
\end{figure}
Due to the blinking of the QD, we observe strong fluctuations between a bright and a dark state for nearly 70~s before photobleaching. A slight drift of the experiment from optimum alignment is responsible for a continuous loss of collected signal with time. By applying a linear correction to account for the drift, we retrieve the distribution of detected events (as shown in \fig{f2}b), where the "on" and the "off" states clearly appear \cite{mahler_towards_2008}. This is a signature of the addressing of a single nanocrystal. Fluorescence decay rate $\Gamma$ of the QD is measured by fitting the decay histogram (see Methods). As expected, $\Gamma$ is greatly enhanced as compared to the decay rate of a single QD on a glass substrate (\fig{f2}c). While the decay of a QD on glass presents a monoexponential decay with a decay rate of 0.034~ns$^{-1}$, the decay statistics of the QD on the nanowire is biexponential, with 80 \% of the total measured photon emission showing a decay rate of 0.67~ns$^{-1}$. The faster contribution shows a very high decay rate (>12~ns$^{-1}$) which we attribute to the formation of biexcitons in the QD. From photon coincidence measurements performed on single QDs within the same experimental conditions, we estimate the biexciton-to-exciton ratio to be on the order of 30~\% \cite{nair_biexciton_2011}, which agrees with the estimate based on the decay histogram (see Supplementary Section \ref{si1} for $g^{(2)}$ photon correlation analysis).

\bigskip

In order to characterize the acceptor fluorescence, we proceeded as follows. Since, in the energy transfer experiment, the acceptor is excited via the plasmon launched by the QD decay, we measured the fluorescence of the acceptor bead when excited by a surface plasmon launched by focusing a pulsed laser at $\lambda = 640$~nm onto the apex of the wire. In order to mimic the QD emission, laser wavelength has been chosen to be close to the QD emission maximum. The decay histogram of the acceptor on glass and on a silver nanowire, while excited via a surface plasmon, is shown in \fig{f2}d. By using a lognormal function to fit the decay histogram (see Supplementary Section \ref{si3}) and the most frequent values of $\Gamma$ as estimate of the decay rate, we measure a Purcell factor $\Gamma/\Gamma_0$ on the order of 2. Interestingly, when the acceptor is coupled to the nanowire, its decay rate is larger when the excitation occurs via the surface plasmon than for a free space excitation (see Supplementary Section \ref{si3} for experimental data). Indeed, in the first situation, only the molecules inside the bead that are well coupled to the nanowire are excited, while in the second situation excitation concerns all molecules inside the bead.   

\bigskip

The occurrence of energy transfer between the QD and the bead implies that the measured acceptor decay histogram will be the convolution of the decay histogram of the donor and acceptor excited independently. The expected signal is shown in the inset of \fig{f2}e (orange curve) and is computed as the convolution of the functions fitting the decay histograms of the donor and the acceptor when excited independently. Experimental data, together with the expected curve, are reported in \fig{f2}e. Their excellent agreement is a proof that the observed acceptor fluorescence comes from energy transfer from the QD via the surface plasmon. Details on signal processing are reported in Supplementary Section \ref{si2}. From the decay histograms, we estimate the total number of fluorescence photon coming from acceptor $N_{2}$ to be of $1010 \pm 50$, while we detect at the same time $N_{1} = 1.6 \; 10^7$ fluorescence photons from the QD. 

\bigskip

Since the acceptor is excited via plasmon energy transfer, we expect the QD and the acceptor to blink simultaneously. This can be proved by characterizing the correlation between the fluorescence intensity $I_1(t)$ and $I_2(t)$ measured for the QD and the acceptor, respectively. Simultaneous blinking will lead to a linear relation, such that $I_2(t) = \alpha \, I_1(t)$ with $\alpha = N_2/N_1$. The degree of correlation, as a function of the delay $\tau$ between the signals detected on each channel, is measured by the coefficient $R(\tau)$ defined as follows:
\begin{equation}
 R(\tau)=\frac{\mathrm{Cov}[I_1(t),I_2(t+\tau)]}{\hat{\sigma}_1 \; \hat{\sigma}_2 }
 \label{e1}
 \end{equation}
where $\mathrm{Cov}[I_1(t),I_2(t+\tau)]$ is the covariance of the two intensities and $\hat{\sigma}_1$ and $\hat{\sigma}_2$ are estimates of the standard deviation for the fluorescence intensities $I_1$ and $I_2$, respectively. With this definition, it follows from the Cauchy-Schwartz inequality that $R$  is equal to $1$ for perfectly correlated variables while it vanishes for uncorrelated variables. In our experiment, the major contribution to the variance of the quantum dot fluorescence is blinking, which dominates on noise. Therefore we can use the standard deviation $\sigma_1$ measured in channel 1 to approximate $\hat{\sigma_1} \approx \sigma_1$ and, since fluorescence of the QD and of the acceptor are linearly dependent, $\hat{\sigma}_2 \approx \alpha \, \sigma_1$ (see Supplementary Section~\ref{si4} for the derivation). In \fig{f3}a (orange curve), we show the correlation coefficient $R(\tau)$ of the QD and the acceptor intensity traces (see Methods). 
\begin{figure}[ht]
\centering
\includegraphics[scale = 0.90]{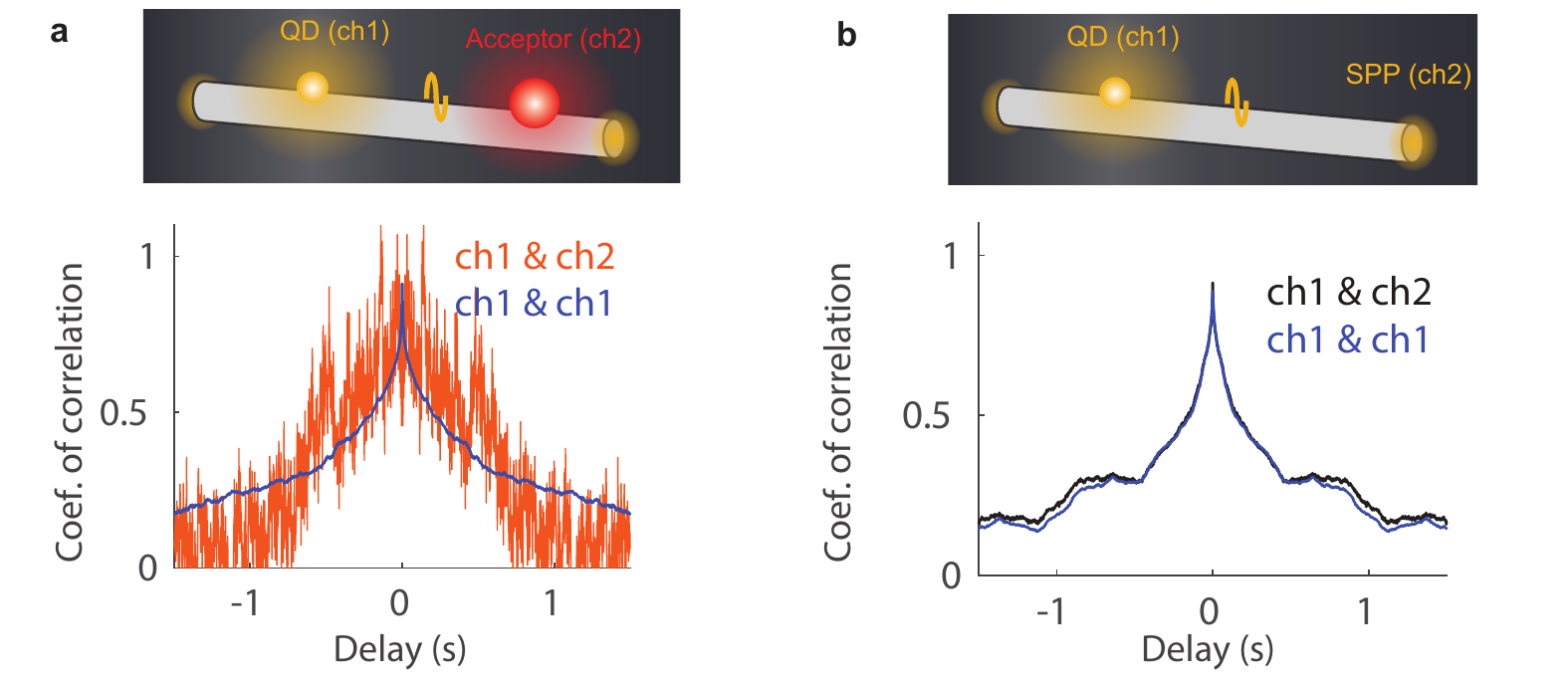} 
\caption{Correlation measurements of the fluorescence of the emitters. \newline \hspace{\textwidth} \small \textbf{a,} Correlation coefficient of the QD fluorescence (channel~1) and the acceptor fluorescence (channel~2) as defined in \eq{e1}. \textbf{b,} Correlation coefficient for an experiment in which we measure the QD fluorescence (channel~1) and the light scattered by the nanowire end (channel~2).}
\label{f3}
\end{figure}
We observe that, while $R$ goes to zero for large delays, the degree of linear correlation between the two intensity traces is almost one for zero delay, proving that the QD and the acceptor blink simultaneously. The small deviation from unity is a consequence of the approximation $\hat{\sigma_1} \approx \sigma_1$. 
In the same figure, we show the correlation coefficient $R(\tau)$ of the intensity trace of the QD with itself (blue curve). The similarity between the two curves proves that the time scales of the fluctuations are effectively the same for both the QD and the acceptor fluorescence. For comparison, we performed an experiment in which a single QD launches a plasmon on a silver nanowire and we detected the light scattered at the nanowire end, located 7~$\mu$m apart. Figure 3b shows the correlation coefficient of the QD fluorescence intensity and the intensity of the scattered at the wire extremity (black curve), as well as the autocorrelation of the QD trace (blue curve). The similarity between the two curves corroborates our analysis. 

\bigskip
%{\blue 
Our work shows that it is possible to control the spontaneous emission of a fluorescent molecule embedded in a nanobead by means of single plasmons launched by a quantum emitter. Indeed, photon emission from the quantum dot and the nanobead are linearly related, which implies that the nanobead exhibits the same blinking behavior of the quantum dot. By enhancing the efficiency of the energy transfer process between the donor and the acceptor, one should observe single photon emission from the nanobead, the signature of which would be the observation of an antibunched emission between the quantum dot and the bead. A significant enhancement of the energy transfer efficiency would be achieved using structures with optimized mode coupling between the donor and the acceptor sites. In this perspective, hybrid plasmon-dielectric waveguides appear as promising structures that would allow an enhancement of the energy transfer efficiency of several order of magnitude \cite{oulton_hybrid_2008} \cite{roque_nanophotonic_2015}. The use of techniques of deterministic nanopositioning of quantum emitters, will allow to further optimize the coupling. Ultimately, efficient coupling of single emitters via plasmonic mode opens new avenues for exciting experiments in the field of quantum plasmonics. %\cite{dzsotjan_quantum_2010}\cite{gonzalez-tudela_entanglement_2011}.
% }
\pagebreak

\section*{Methods}
\paragraph*{Sample preparation} On a thin glass coverslip cleaned with a UV/ozone surface cleaner, we spin-coat for 30~s a dilute solution of silver nanowires $\sim$115 nm in diameter (Sigma Aldrich) and polystyrene fluorescent beads $\sim$176 nm in diameter (FluoSpheres Microsphere Dark Red, ThermoFisher Scientific) in isopropyl alcohol. After a few minutes, we spin-coat for 30~s a dilute solution of chemically synthesized CdSe/CdS/ZnS quantum dots in hexane (7.5 $\pm$ 0.5~nm in diameter, measured with transmission electron microscopy images). The resulting sample presents several nanowires with different lengths ranging from 5 to 50 $\mu$m as well as isolated QDs and fluorescent beads, some of them in near-field coupling with a nanowire.

\paragraph*{Optical setup} We excite the quantum dots with a pulsed laser diode emitting at 405~nm (LDH Series P-C-405M, PicoQuant). The average power on the sample is $\sim$~2~$\mu$W. For sample excitation and fluorescence collection, we use an oil immersion microscope objective (magnification $\times100$, N.A. $= 1.4$, Olympus). A dichroic mirror (Di01-R488, Semrock) and a long-pass filter (BLP01-488R, Semrock) selects fluorescence photons. By passing through a 50/50 beamsplitter, fluorescent photons are detected by an electron multiplying CCD camera (iXon3, Andor) and by two avalanche photodiodes (PDM-R, Micro Photon Devices). 
The avalanche photodiodes are mounted in HBT configuration by using either a 50/50 beamsplitter or a dichroic mirror (ZT647rdc, Chroma) together with a long-pass filter (ET665lp, Chroma). The signal collected by the photodiodes arises from a 450 nm sized region of the sample which is optically conjugated with the detection system. A removable mirror can be inserted in the optical path to direct photons towards a fibered spectrometer (Acton SP2300, Princeton Instruments). We use a piezoelectric positioning system (PXY 200SG, Piezosystem Jena) to perform lateral displacements of the sample in the object plane of the microscope objective.

\paragraph*{Determination of the distance between the emitters}

We measure distances in the sample plane from images captured with the EM-CCD camera. Each image is the sum of 200 acquisitions, each of these with a gain of 80 and an integration time of 0.1~s. By using the piezoelectric positioning system, we find the pixel size reported on the sample to be 169~nm. Then, we point bright spots in the image and locate them using a two dimensional Gaussian fit with, as free parameters, the hight, the center position and the width. This procedure gives a resolution of approximatively 50~nm. 

\paragraph*{Decay rate analysis}

We measure the decay rate of fluorescent emitters by fitting the fluorescence histogram with the convolution of the instrument response function (IRF) and a model function. The decay histogram of isolated QDs and isolated beads on glass is well fitted by a mono-exponential function with two free parameters, the amplitude and the decay rate. The decay rate of a single QD coupled to a silver nanowire is well fitted by a bi-exponential function (4 free parameters, the amplitude and the decay rate for each exponential decay). The fast decay rate contribution is around 20 \% of the total emission. The decay rate of a fluorescent bead coupled to a silver nanowire is well fitted by a lognormal distribution of decay rates. This fit function has 3 free parameters: the amplitude $A$, the most frequent decay rate $\Gamma_{mf}$ and a parameter characterizing the width of the distribution $w$. More details can be found in Supplementary Section \ref{si3}. 

\paragraph*{Covariance coefficient calculation}

We calculate the covariance coefficient $R(\tau)$ of the centred values of the intensity traces measured on channel 1 and 2, with a time resolution of 1~ms. In order to have a large variance in the signals, we select a 30~s time interval (from $t$=40 to 70~s in \fig{f2}a) during which the QD blinks frequently. We symmetrize the intensity time traces to be insensitive to any drift of the experiment. Furthermore, in order to improve the signal to noise ratio, we select the photons on the base of their arrival time with respect to the laser excitation pulse. For channel~2 (acceptor fluorescence), we select photons detected between 1.856~ns and 12~ns, where the signal/noise ratio is of the order of unity.

\bibliographystyle{naturemag}
%\bibliography{references}

\section*{Acknowledgements}
The authors thank N. Lequeux and T. Pons for helping in sample preparation. This work was supported by LABEX WIFI (Laboratory of Excellence ANR-10-LABX-24) within the French Program Investments for the Future under reference ANR-10- IDEX-0001-02 PSL*, by the Region Ile-de-France in the framework of DIM Nano-K, by the Programme Emergences 2015 of the City of Paris, by PSL Research University in the framework of the project COSINE and by ANR Nanodose.
\section*{Author contributions}
Experiments and data analysis were carried out by D. B. under the supervision of V. K. and Y. D. W, using avalanche photodiodes developed by A. G. and I. R.. Semiconductor quantum dots were synthesized by E. L. and S. I.. The manuscript was written by D. B. and V.K. with contributions from all co-authors.
\section*{Competing financial interests}
The authors declare no competing financial interests.

%%%%%%%%%%%%%%%%%%%%%%%%%%%%%%%%%%%%%%%%%%%%%%%%%%%%%%%%%%%%%%%%%%%%%%%
\pagebreak
\setcounter{equation}{0}
\setcounter{figure}{0}

\section*{Supplementary information}

\subsection{Photon coincidence measurements of single quantum dots}
\label{si1}

\begin{figure}[H]
\centering
\includegraphics[scale = 0.75]{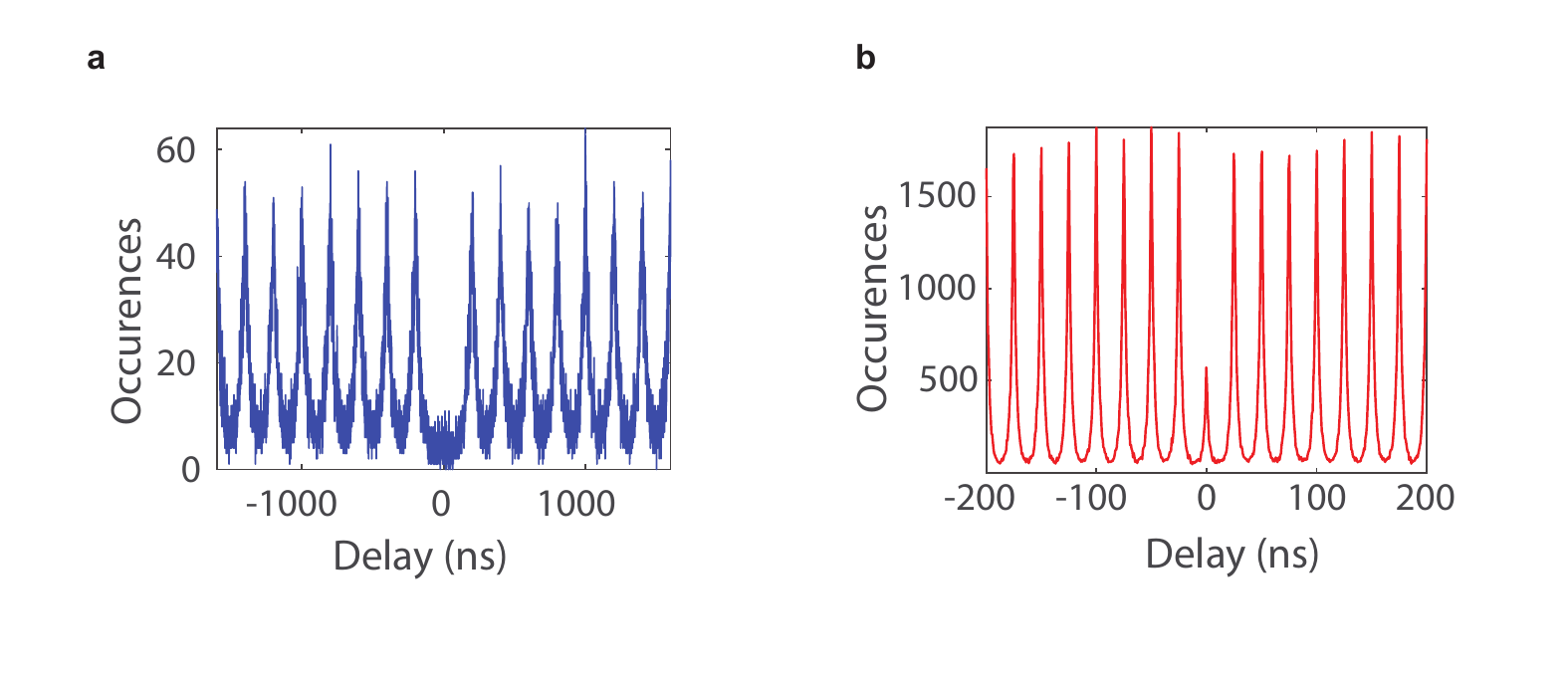}
\caption{Photon coincidence measurements. \newline \hspace{\textwidth} \small \textbf{a,} Single CdSe/CdS/ZnS nanocrystal on glass. Repetition rate: 5~MHz. Measured fluorescent counts of the bright state: 20~c/ms. \textbf{b,} Single CdSe/CdS/ZnS nanocrystal on silver nanowire. Repetition rate: 40~MHz. Measured fluorescent counts of the bright state: 300~c/ms.}
\label{fs1}
\end{figure}

In this section we present experimental data of photon coincidence measurements performed on single CdSe/CdS/ZnS quantum dots. For a single nanocrystal on a glass coverslip excited with an average laser power of $\sim$~0.5~$\mu$W (repetition rate 5~MHz, pulse duration 600~ps) we measure a strong antibunching (\fig{fs1}a). When increasing the excitation power to the one used to perform the measurements reported in the manuscript (average power of $\sim$~2~$\mu$W, repetition rate 40~MHz, pulse duration 600~ps), the second order correlation function shows a small peak at zero delay, characteristic of the formation of biexcition (\fig{fs1}b). By measuring the ratio between the area under the zero delay peak and the adjacent ones, we find a biexciton-to-exciton ratio of 4~\% and 29~\%, respectively.

\subsection{Energy transfer decay histogram analysis}
\label{si2}

\begin{figure}[H]
\centering
\includegraphics[scale = 0.9]{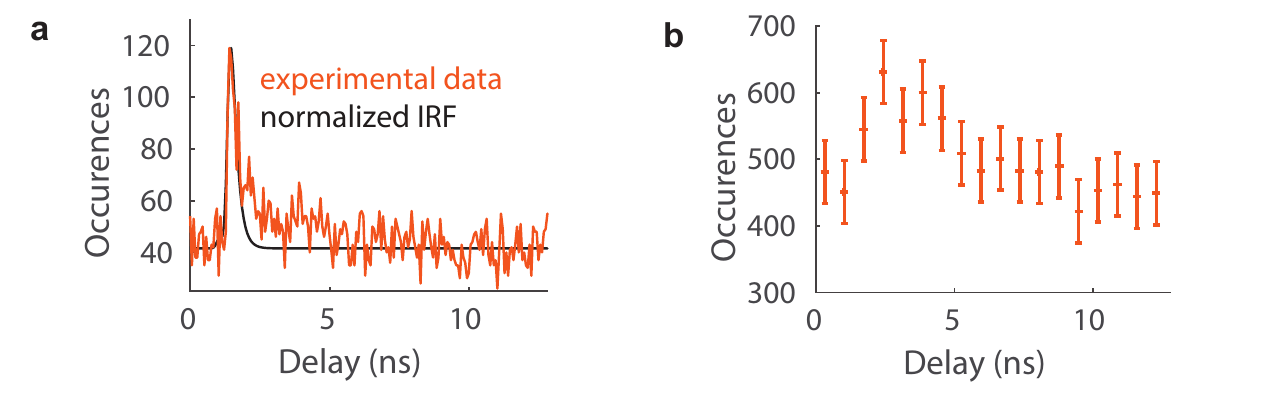}
\caption{Decay histogram measured on channel 2. \newline \hspace{\textwidth} \small \textbf{a,} Experimental data binned at 64~ps (in orange). The black curve is a fit of the data by the IRF.  \textbf{b,} Corrected data binned at 700~ps.}
\label{fs2}
\end{figure}

In this section we present raw acceptor decay histogram and the data processing that has been made on it in order to prove the occurrence of the energy transfer. Figure \ref{fs2}a presents raw data, measured with a 64~ps resolution. The decay histogram shows two components. The first one is characterized by a long-lifetime, characteristic of the occurrence of energy transfer. The second one is characterized by a very short lifetime which is well fitted by the instrument response function of the experimental setup. Such fast component of the decay histogram has been observed in a test experiment, performed on bare silver nanowires deposited on a glass coverslip under pulsed excitation at $\lambda$~=~405~nm and is attributed to silver luminescence. Luminescence is characterized by a very broad spectrum and can propagate along the nanowire via a surface plasmon, for finally being scattered by the acceptor bead and therefore detected. 

We removed this contribution by fitting the instrument response function to the data and by subtracting it from the decay histogram. Figure \ref{fs2}b shows the corrected decay histogram with a 700~ps resolution. As explained in the main text, error bars are equal to the square root of the total number of detected events.

\subsection{Decay rate of the acceptor bead coupled to a silver nanowire}
\label{si3}

\begin{figure}[H]
\centering
\includegraphics[scale = 0.95]{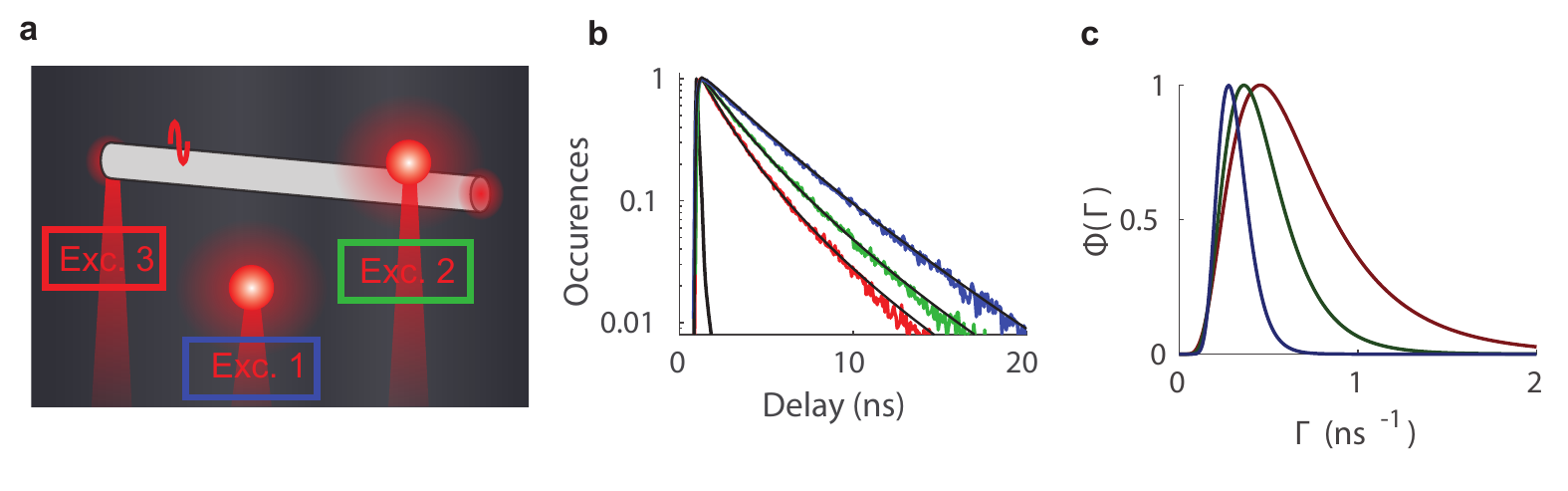}
\caption{Measurement of the acceptor bead decay rate. \newline \hspace{\textwidth} \small \textbf{a,} Scheme of the experiment. \textbf{b,} Measured decay histogram in three different situations: bead not coupled to a nanowire (blue), bead coupled to a nanowire and excited directly by the laser (green), same bead excited by a surface plasmon launched by focusing a laser on the extremity of the nanowire (red). \textbf{c,} Distribution of decay rates for these three situations.}
\label{fs3}
\end{figure}

In this section we present the measurement of the decay rate of the acceptor bead when the excitation does not occur via energy transfer. 
The different excitations performed are schematically reported in \fig{fs3}a and the corresponding decay histograms are reported in \fig{fs3}b following the same colour scheme as in \fig{fs3}a. As a reference, we measured the decay histogram of the acceptor on a glass coverslip (blue curve). Then we compared it to the decay histogram obtained for the acceptor bead coupled to a silver nanowire in two different situations. The bead can either be excited by a laser ($\lambda=642$~nm) focused on it (green curve) or by a surface plasmon propagating on the silver wire (red curve). The plasmon is launched by focusing a laser ($\lambda=642$~nm) on the extremity of the wire.      

As expected, the decay rate of the bead coupled to the wire (for both excitation schemes) is larger than on glass. Moreover, the decay rate obtained when the bead is excited via the surface plasmon is larger than the one obtained with a far field excitation. Indeed, in the first situation, only the acceptor molecules located closer to the wire are excited, while in the second situation all the molecules in the bead are excited. 
In order to get a quantitative measurement of the decay rate, we fit the decay histograms with the convolution of the instrument response function (IRF) and a lognormal distribution of decay rates which reads:
$$
I_{log}(t) = A \int_{\Gamma=0}^{\infty} \Phi(\Gamma) \exp(-\Gamma t) \mathrm{d} \Gamma \qquad \mathrm{with} \qquad \Phi(\Gamma) = \exp \left(-\frac{\ln^2(\Gamma/\Gamma_{mf})}{w^2} \right).
$$
In the formula above, $A$ is the amplitude, $\Gamma_{mf}$ the most frequent decay rate and $w$ a parameter characterizing the width of the distribution. Figure \ref{fs3}c shows the measured distribution of decay rates $\Phi(\Gamma)$ for the three situations depicted in \fig{fs3}a. 

\subsection{Characterization of the degree of linear correlation between intensity time traces }
\label{si4}

In this section we characterise the degree of linear correlation between two intensity time traces $I_1(t)$ and $I_2(t)$. Each time trace $i$ can be written as a sum of the fluorescence counts $I'_i(t)$ and the noise counts $I''_i(t)$, reading
\begin{equation}
I_1(t) = I'_1(t) + I''_1(t)
\end{equation}
\begin{equation}
I_2(t) = I'_2(t) + I''_2(t)
\end{equation}
We define $\overline{I_1}$ and $\overline{I_2}$ as the mean value on the time interval $\Delta t$ of, respectively, $I_1(t)$ and $I_2(t)$. Similarely, we write
\begin{equation}
\overline{I_1} = \overline{I'_1}+ \overline{I''_1}
\end{equation}
\begin{equation}
\overline{I_2} = \overline{I'_2}+\overline{I''_2}
\end{equation}
On this time interval $\Delta t$, the covariance of the intensity time traces reads
\begin{equation}
\mathrm{Cov}[I_1(t),I_2(t)] = \int_{\Delta t}(I_1(t)-\overline{I_1})( I_2(t)-\overline{I_2}) \; \mathrm{d}t
\end{equation}
It follows
\begin{multline}
\mathrm{Cov}[I_1(t),I_2(t)]   = \int_{\Delta t}(I'_1(t)-\overline{I'_1})( I'_2(t)-\overline{I'_2}) \; \mathrm{d}t  + \int_{\Delta t}(I'_1(t)-\overline{I'_1})( I''_2(t)-\overline{I''_2}) \; \mathrm{d}t  \\
+ \int_{\Delta t}(I''_1(t)-\overline{I''_1})( I'_2(t)-\overline{I'_2}) \; \mathrm{d}t  + \int_{\Delta t}(I''_1(t)-\overline{I''_1})( I''_2(t)-\overline{I''_2}) \; \mathrm{d}t 
\end{multline}
The last three contributions involve the integral over $\Delta t$ of a white noise and thus cancel out. The covariance is therefore determined by the fluorescence counts only, and reads 
\begin{equation}
\mathrm{Cov}[I_1(t),I_2(t)]   = \int_{\Delta t}(I'_1(t)-\overline{I'_1})( I'_2(t)-\overline{I'_2}) \; \mathrm{d}t 
\end{equation}
This equation equals zero if $I'_1(t)$ and $I'_2(t)$ are uncorrelated. However, if a linear relation between $I'_1(t)$ and $I'_2(t)$ exists such that $I'_2(t) = \alpha \; I'_1(t)$, this equation reads
\begin{equation}
\mathrm{Cov}[I_1(t),I_2(t)]  = \alpha \times \int_{\Delta t}(I'_1(t)-\overline{I'_1})^2 \; \mathrm{d}t 
\end{equation}
We define the correlation coefficient $R$ as follows:
\begin{equation}
 R=\frac{\mathrm{Cov}[I_1(t),I_2(t)] }{\alpha \times \int_{\Delta t}(I'_1(t)-\overline{I'_1})^2 \; \mathrm{d}t} 
 \end{equation} 
We define  $\sigma'_1$ and $\sigma'_2$ as the standard deviation on the time interval $\Delta t$ of the fluorescence contribution of each trace, namely, $I'_1(t)$ and $I'_2(t)$. They read: 
\begin{equation}
\sigma'_1 = \sqrt{\int_{\Delta t}(I'_1(t)-\overline{I'_1})^2 \; \mathrm{d}t}
\end{equation}
\begin{equation}
\sigma'_2 = \sqrt{\int_{\Delta t}(I'_2(t)-\overline{I'_2})^2 \; \mathrm{d}t}
\end{equation}
From $I'_2(t) = \alpha \, I'_1(t)$, we can directly write $\sigma'_1 = \alpha \, \sigma'_2$. With these notations, the correlation coefficient $R$ reads
\begin{equation}
 R=\frac{\mathrm{Cov}[I_1(t),I_2(t)] }{\sigma'_1 \, \sigma'_2} 
 \end{equation} 
Using this definition, 
\begin{itemize}
\item if $I'_1(t)$ and $I'_2(t)$ are uncorrelated then $R=0$,
\item if $I'_1(t)$ and $I'_2(t)$ are linearly related then $R =1$. 
\end{itemize} 

\subsection{Synthesis of CdS/CdSe/ZnS nanocrystals}
\label{si5}
\subsubsection{Chemicals}
1-Octadecene (ODE, 90 \%, Aldrich), oleylamine (70 \%, Fluka), oleic acid (90 \%, Aldrich), sodium myristate (99 \%, Fluka), cadmium nitrate (99.999 \%, Aldrich), cadmium oxide (99.99 \%, Aldrich), zinc nitrate (aldrich, 98 \%), Sulfur powder (Aldrich, 99;998 \%), selenium powder 100 mesh (99.99 \%, Aldrich), sulfur (99.998 \%, Aldrich), ethanol (Carlo Erba, 99.5 \%), methanol (VWR, 100 \%), n methyl formamide (NMFA) (Aldrich, 99 \%). 

\subsubsection{Precursor preparation}

${\bf Cd(Myr)}_2$: 3.2g (80 mmol) of NaOH are dissolved in 500 mL of methanol. Then 18.2~g of myristic acid are added in the flask. The whitish solution is stirred for 15~min. Meanwhile, 8.2~g  of cadmium nitrate tetrahydrate are dissolved in 50~mL of methanol. This solution is added to the sodium myristate solution and  a white precipitate gets formed. The solution is further stirred for 15~min. The white solid is isolated by filtration, and washed several times with methanol. The solid is finally dried under vacuum overnight.
\\
{\bf S-ODE~0.1M:}  150~mL of octadecene are degassed under vacuum for 30~min. Then the flask is put under Ar and 480~mg of S powder are introduced in the flaks. The solution is heated at $140\degree C$ until the formation of a yellow clear solution. The flask is finally cooled down.
\\
{\bf Se-ODE 0.1M:}  140~mL of octadecene are degassed under vacuum for 30~min. Then the flask is put under Ar and the tempearture raised at $170\degree C$. 1.18~g of Se powder is mixed in 10~mL of octadecene and introduced in the flask dropwise. The temperature is finally raised to $205 \degree C$ and the flask heated for 30~min. The final solution is clear and yellow-orange. The flaks is finally cooled down.
\\
${\bf Cd(OA)_2 0.5M}$: In a 250~ml three-neck flask, 6.42~g of CdO and 100~mL of oleic acid are introduced and the mixture is heated at $180\degree C$ until the formation of a yellow clear solution. Then the flask is cooled below $120\degree C$ and put under vacuum for 30~min. The flaks is finally cooled down.
\subsubsection{QD synthesis}

In a three neck flask we introduce 170~mg of cadmium myristate with 7.5~mL of octadecene. The flask is degassed under vacuum at room temperature for half an hour. The atmosphere is then switched to Ar and the temperature raised to $250\degree C$. 12~mg of selenium powder are mixed with 1~mL of ODE and sonicated, before being quickly injected into the flask. After 5~min at $250\degree C$, 0.2~mL of oleic acid  and 2~mL of oleylamine are added. Then a mixture of 1~mL cadmium oleate (0.5M) and 5~mL of Se-ODE (0.1M) is prepared and half of them is injected with a 5~mL/h flowrate. The reaction is then cooled down to room temperature and the nanocrystal precipitated by addition of ethanol. After centrifugation the QD pellet is redispersed in 10~mL of octadecene.

For the growth of the CdS shell, A SILAR method is used, where we successively introduce precursor for Cd and S. 4~ml of the previous solution of CdSe core are mixed with 16~mL of octadecene and 10~mL of oleylamine. The solution is degassed for 30~min. Then cadmium oleate (0.1M) and S-ODE (0.1M) are introduced in the flask and the temperature raised at $230\degree C$. After 20~min, cadmium oleate (0.1M) is added and after 10~min S-ODE (0.1M) is added. We repeat this addition of Cd and S up to the formation of four layer of CdS. The reaction is then cooled down to room temperature and the nanocrystal precipitated by addition of ethanol. After centrifugation the QD pellet is redispersed in 10~mL of octadecene.

For the growth of ZnS shell use the C-ALd procedure as developed in \cite{Ithurria2012}. Briefly a 0.2M solution of Na2S in N methyl formamide (NMF) and a 0.2M solution of $\mathrm{Zn(NO_3)_2.6H_2O}$ in NMF are prepared. The CdSe/CdS QD dispersed in hexane are mixed with the same amount of NMF. Some of the NA2S solution is added and the solution is stirred until we observe a phase transfer. The particle are then  precipitated by addition of ethanol and centrifuge. The formed pellet is redispersed in fresh NMF. Some of the zinc solution is added and the same cleaning procedure is repeated. The growth of a second ZnS layer is conducted. The particles are precipitated and cleaned again and an excess of oleic acid is added to the vial to retransfer the QD toward a non polar phase.

\bibliographystyle{naturemag}
%\bibliography{references}

\end{document}